\documentclass{ws-procs9x6}

\usepackage{wasysym}

\def\de{\delta}

\def\ka{\kappa}
\def\la{\lambda}

\def\si{\sigma}

\def\ph{\phi}

\def\om{\omega}

\def\Om{\Omega}

\def\mn{{\mu\nu}}

\def\fr#1#2{{{#1} \over {#2}}}

\def\frac#1#2{{\textstyle{{#1}\over {#2}}}}

\def\lsim{\mathrel{\rlap{\lower4pt\hbox{\hskip1pt$\sim$}}
    \raise1pt\hbox{$<$}}}
\def\gsim{\mathrel{\rlap{\lower4pt\hbox{\hskip1pt$\sim$}}
    \raise1pt\hbox{$>$}}}

\def\part2{\partial_\alpha \partial^\alpha}

\def\etal{{\it et al.}}

\def\tt{$t^{\ka\la\mu\nu}$}

\def\sss{s^{\mu\nu}}
\def\ttt{t^{\ka\la\mu\nu}}

\def\sb{\overline{s}}
\def\tb{\overline{t}}
\def\ub{\overline{u}}

\def\xx'{|\vec x -\vec x'|}

\newcommand{\beq}{\begin{equation}}
\newcommand{\eeq}{\end{equation}}
\newcommand{\bea}{\begin{eqnarray}}
\newcommand{\eea}{\end{eqnarray}}
\newcommand{\bit}{\begin{itemize}}
\newcommand{\eit}{\end{itemize}}
\newcommand{\rf}[1]{(\ref{#1})}

\begin{document}

\title{Testing Lorentz Symmetry with Gravity}

\author{Quentin G.\ Bailey}

\address{Physics Department \\
Embry-Riddle Aeronautical University\\
3700 Willow Creek Road  \\ 
Prescott, AZ 86301, USA \\ 
E-mail: baileyq@erau.edu}

\maketitle

\abstracts{
In this talk, 
results from the gravitational sector of the Standard-Model Extension 
(SME) are discussed.  
The weak-field phenomenology of the resulting modified 
gravitational field equations is explored.
The application of the results to a variety of 
modern gravity experiments, 
including lunar laser ranging, Gravity Probe B, 
binary pulsars, and Earth-laboratory tests, 
shows promising sensitivity to gravitational coefficients 
for Lorentz violation in the SME.}

\section{Introduction}
\label{Introduction}

At the present time, 
a comprehensive and successful description of nature is 
provided by general relativity and the Standard Model of 
particle physics.
It is expected, 
however,
that a single underlying unified theory  
would merge them at the Planck scale.
To date, 
a completely satisfactory theory remains elusive.
Experimental clues about this underlying theory are 
lacking since direct measurements at the Planck scale 
are infeasible at present.

An alternative approach is to 
look for suppressed new physics effects
coming from the underlying theory 
that are potentially detectable in modern sensitive experiments. 
One promising class of signals satisfying this criteria are 
minuscule violations of Lorentz symmetry.\cite{cpt04} 
For describing the observable signals of Lorentz violation,
the effective field theory known as the Standard-Model 
Extension (SME) provides a useful tool.\cite{kpck,akgrav}

Much of the theoretical and experimental work
on the SME has involved the the Minkowski-spacetime limit.
Experimental studies have included ones
with photons\cite{photonexpt},
electrons\cite{eexpt},
protons and neutrons\cite{ccexpt},
mesons\cite{hadronexpt},
muons\cite{muexpt},
neutrinos\cite{nuexpt},
and the Higgs.\cite{higgs}
Though no compelling evidence for Lorentz violation has been found,
only about half of the possible signals
involving light and ordinary matter have been experimentally 
investigated,
while some other sectors remain largely unexplored.
The subject of the talk will be a recent SME-based study of 
gravitational experiments searching for violations of 
local Lorentz invariance.
For a more detailed discussion, 
the reader is referred to Ref.\ \refcite{qkgrav}.

\section{Theory}
\label{theory}

The gravitational couplings in the SME action are presented
in Ref.\ \refcite{akgrav}.
The geometric framework assumed is a Riemann-Cartan spacetime,
allowing for torsion.
For simplicity, 
attention is restricted to the Riemann-spacetime limit.
In this limit, 
the effective action of the pure-gravity minimal SME is written
\beq
S = \fr {1}{16\pi G} \int d^4x \sqrt{-g}[(1-u)R + \sss R^T_\mn 
+ \ttt C_{\ka\la\mu\nu}]+ S^\prime.
\label{act}
\eeq
Here $R$ is the Ricci scalar, 
$R^T_\mn$ is the trace-free Ricci tensor, 
and $C_{\ka\la\mu\nu}$ is the Weyl conformal tensor.  
The leading Lorentz-violating gravitational couplings
are controlled by the coefficients for Lorentz violation 
$u$, $s^\mn$, and $t^{\ka\la\mu\nu}$.
Equation \rf{act} contains 20 independent coefficients,
of which 1 is in $u$,
9 are in the traceless $s^\mn$, 
and 10 are in the totally traceless $t^{\ka\la\mu\nu}$.

It is known that explicit Lorentz violation,
whereupon the coefficients for Lorentz violation in Eq.\ \rf{act} 
are nondynamical functions of spacetime,
is generally incompatible with Riemann spacetime.\cite{akgrav}
Spontaneous Lorentz violation, 
however, 
evades this problem\cite{bkgrav} and is the approach
adopted to analyze Eq.\ \rf{act}.
In this scenario the coefficients 
$u$, 
$s^\mn$, 
and $t^{\ka\la\mu\nu}$ are dynamical fields 
that acquire vacuum expectation values denoted
$\ub$, 
$\sb^\mn$, and $\tb^{\ka\la\mu\nu}$. 
The general matter action $S^\prime$ in Eq.\ \rf{act} 
therefore includes the dynamics for ordinary matter as
well as the coefficients for Lorentz violation.

To construct the field equations associated with the action
\rf{act}, 
while taking into account the unknown dynamics
of the coefficient fields $u$, 
$s^\mn$, 
and $t^{\ka\la\mu\nu}$, 
represents a challenging theoretical task.
In the case of weak-field gravity, 
however, 
a set of modified field equations can be obtained under
mild assumptions,\cite{qkgrav} which then
determine the leading corrections to general relativity 
arising from Lorentz violation. 
In particular, 
the dominant terms in the post-newtonian metric can be determined.
From the post-newtonian metric 
an effective classical lagrangian for $N$ point-like 
bodies can be derived.
This lagrangian provides the basis for studies of orbital 
experiments probing the coefficients $\sb^\mn$, 
while the post-newtonian metric is used to describe experiments
probing spacetime geometry. 

It is standard to compare a given 
post-newtonian metric with the Parametrized Post-Newtonian
(PPN) metric.\cite{cmw,ppn}
It turns out that the match can only be achieved when 
the SME coefficients $\sb^\mn$ are assumed isotropic
in a special coordinate frame, 
resulting in only one rotational scalar coefficient 
(taken as $\sb^{00}=\sb^{jj}$) remaining. 
This isotropic assumption is not generally adopted in SME
studies and so the relationship between the SME and the PPN
is one of partial overlap. 

\section{Lunar laser ranging}
\label{llr}

The primary observable in lunar laser ranging experiments
are oscillations in the Earth-Moon distance.
High sensitivity is achieved by timing laser pulses reflecting
off of one or more of the five reflectors on the lunar 
surface.\cite{llrrev}
Appropriate application of the effective classical lagrangian 
yields the Lorentz-violating corrections to the Earth-Moon coordinate 
acceleration.
Ideally, 
a computer code would be used that includes the standard 
dynamics of the Earth-Moon system and effects from the pure-gravity 
sector of the minimal SME.

It is useful, 
however,
to perform a perturbative analysis
that extracts the dominant oscillation frequencies 
and corresponding amplitudes for Earth-Moon separation
oscillations driven by Lorentz violation.
The radial corrections $\de r$ 
arising from the Lorentz-violating terms in the acceleration 
take the generic form
\beq
\de r = \sum_n [A_n \cos (\om_n T + \phi_n) 
+ B_n \sin (\om_n T + \phi_n)].
\label{delr}
\eeq
The dominant amplitudes are denoted $A_n$ and $B_n$ 
and the corresponding phases are $\ph_n$.
For example, 
one oscillation occurs at twice the mean orbital frequency
$\om$ with amplitudes given by
$A_{2\om} = -\fr {1}{12}(\sb^{11}-\sb^{22}) r_0 $ and 
$B_{2\om} = -\fr 16 \sb^{12} r_0 $ where $r_0$ is the
mean Earth-Moon distance.
The coefficients $\sb^{11}-\sb^{22}$ and $\sb^{12}$ are combinations
of the standard Sun-centered frame coefficients $\sb^{JK}$,
and depend on them through angles describing the orbit. 
This angular dependence indicates that it 
may be useful to consider artificial satellite orbits of 
varying orientation, 
in order to attain sensitivity to coefficients that may elude 
the lunar orbit.

For lunar laser ranging,
at least 5 independent combinations 
of coefficients for Lorentz violation can be measured.
Using standard lunar values and assuming 
ranging precision at the centimeter level,\cite{llrrev} 
the estimated experimental sensitivities are parts in $10^{10}$ 
on combinations of coefficients in $\sb^{JK}$
and parts in $10^{7}$ on the coefficients $\sb^{TJ}$.
An analysis studying the dominant Earth-Moon 
oscillations using $30+$ years of data 
has recently been performed and has achieved roughly this level
of sensitivity.\cite{jb} 
The new Apache Point Observatory Lunar Laser-ranging Operation (APOLLO),
may substantially improve these sensitivities.\cite{apollo}

\section{Gyroscope experiment}
\label{gyroscope experiment}

In general relativity there are two well-known
types of precession of the spin of a freely falling test body in 
the presence of a massive spinning body like the Earth.\cite{lis} 
These two types of spin precession are the geodetic precession
about an axis perpendicular to the body's orbit 
and the gravitomagnetic precession about the spin axis 
of the Earth.
In the context of the pure-gravity sector
of the minimal SME there is an additional precession effect
that occurs due to Lorentz violation.

Ultimately the dominant measurable effects controlled
by the SME coefficients reveal themselves in
the secular evolution of the gyroscope spin $\vec S$, 
described by $d\vec S/dt = g v_0 \vec \Om \times \vec S$
where $g=GM_{\oplus}/r_0^2$ is the mean value of the
gravitational acceleration at the orbital radius $r_0$ 
and $v_0$ is the mean orbital velocity.
The precession vector $\vec \Om$ 
is split into two pieces via
$\Om^J = \Om^J_E + \Om^J_{\sb}$,
with the first term containing precession 
due to conventional effects in general relativity,
and the second term containing contributions from 
the coefficients for Lorentz violation.
The latter is given by
\bea
\Om^J_{\sb} &=& 
\frac 98 (\sb^{TT} 
-\sb^{KL} \hat \si^K \hat \si^L) \hat \si^J
+ \frac 54 \sb^{JK} \hat \si^K,
\label{oms}
\eea
where the result is written in the Sun-centered frame, 
$\hat \si$ is a unit vector normal to the orbital plane,
and contributions from the Earth's inertia have been suppressed.

The result \rf{oms} gives contributions to the 
precession about the orbital angular momentum 
axis $\hat \si$ and the Earth's spin axis $\hat J$.
In addition, 
however, 
there is a qualitatively new precession about the axis defined by 
$\hat n = \hat \si \times \hat J$, 
that is due entirely to Lorentz violation controlled by
the $\sb^{JK}$ coefficients.
Data from the Gravity Probe B (GPB) experiment 
could potentially measure the combinations of 
coefficients occurring in Eq.\ \rf{oms}.\cite{gpb}
If the spin precession vector in three orthogonal directions
can be extracted,
including the $\hat n$ direction,
then attainable sensitivity to $\sb^{JK}$ coefficients
is expected to be at the $10^{-4}$ level, 
given GPB projected sensitivities. 

\section{Binary pulsars}
\label{binary pulsars}

A particularly useful testing ground for general relativity 
is the binary-pulsar system.\cite{bp1,cmw} 
In particular, 
such systems contain compact objects
and high orbital velocities which make them appropriate for 
studies of strong-field gravity.
Pulsar timing data from binary pulsar systems also 
offers the possibility of probing SME coefficients 
for Lorentz violation. 

The Einstein-Infeld-Hoffman (EIH) lagrangian 
describes the post-newtonian dynamics of such systems 
and represents a standard approach.\cite{cmw,mk}
To obtain the key features arising from Lorentz violation,
however,
a point-mass approximation suffices and appropriate use can be 
made of the effective classical lagrangian.
The basic orbit can be modeled as a perturbed 
elliptic two-body problem, 
where six standard orbital elements are used to describe 
the orbit: $a$, 
$e$,
$l_0$,
$i$,
$\Om$,
and $\om$.
Ultimately a pulsar timing formula is used to model 
the number of pulses received as a function of 
arrival time.
The timing formula receives modifications 
due to Lorentz violation from two sources.
First,
the orbital elements, 
with the exception of $a$, 
acquire secular Lorentz-violating corrections.
Second, 
the timing formula itself involves 
an explicit dependence on combinations of 
coefficients for Lorentz violation.

Some simple estimates of sensitivities
reveal that, 
for example,
data from the binary pulsar system PSR $1913+16$
could yield sensitivities\cite{bp2} to Lorentz violation 
at the level of $\sb_e \lsim 10^{-9}$ and $\sb_{\om} \lsim 10^{-11}$
where $\sb_e$ and $\sb_\om$ are the combinations
of coefficients relevant for the orbital elements $e$ and $\om$. 
These combinations of coefficients will also 
change with the orientation of the binary pulsar system.

\section{Other tests}
\label{other tests}

Other types of gravitational experiments
have been explored for their merits in
probing the SME coefficients $\sb^\mn$.\footnote{Analysis
of the classic tests is in Ref.\ \refcite{qkgrav}.} 
In particular, 
Earth-laboratory tests studying gravitational 
interactions between either two controlled masses
or between a test body and the Earth could be used.

One prediction is a newtonian potential 
between two point masses that is modulated
by an anisotropic term $\hat x^{\hat j} \hat x^{\hat k} \sb^{\hat j\hat k}$,
where the unit vector $\hat x$ points between the two masses.
The SME coefficients $\sb^{\hat j\hat k}$ are taken in
the Earth-laboratory frame of reference.
Although the inverse-distance behavior 
of the usual newtonian potential
is maintained, 
the associated force is generally misaligned 
relative to the unit vector $\hat x$.
It is conceivable that experiments studying short-range
tests of gravity might be used to probe these 
coefficients.\cite{isql}
Currently, 
an analysis of this type is underway.\cite{jl}

When considering the effects of the Earth's gravity
on test bodies near the surface of the Earth,
a modified local gravitational acceleration arises.
In the local laboratory frame of reference this
acceleration has a vertical ($\hat z$) component
which is time dependent on sidereal day and year time scales.
Experiments with gravimeters are ideally suited 
for probing such a time variation.\cite{cmw}
An analysis using gravimeter data to extract 
measurements on combinations of coefficients
occurring in this modified acceleration is currently 
underway.\cite{hm}
In addition,  
the local acceleration in the horizontal
directions $\hat x$ (south) and $\hat y$ (east) 
receives modifications from Lorentz violation.

\end{document}